# Bayesian modeling of two-species bacterial competition growth and decline rates in milk

E.J. Quinto[a]*, J.M. Marín[b], I. Caro[a,c], J. Mateo[c], D.W. Schaffner[d]

[a] Department of Nutrition and Food Science, College of Medicine, University of Valladolid, 47005, Valladolid, Spain.

[b] Department of Statistics, University Carlos III de Madrid, 28903 Getafe, Madrid, Spain.

[c] Department of Food Hygiene and Food Technology, University of León, Campus de Vegazana s/n, 24071 León, Spain.

[d] Department of Food Science, Rutgers University, New Brunswick, NJ 08901, USA.

*Corresponding author: E. J. Quinto, Department of Nutrition and Food Science, College of Medicine, Avda Ramon y Cajal 7, University of Valladolid, 47005, Valladolid, Spain. Telephone: +34-983-184943. Fax: +34-983-423812. E-mail: equinto@ped.uva.es.

E-mail addresses: equinto@ped.uva.es (E. J. Quinto), jmmarin@est-econ.uc3m.es (J. M. Marín), icarc@unileon.es (I. Caro), jmato@unileon.es (J. Mateo), don.schaffner@rutgers.edu (D. W. Schaffner).

**Abstract**

Shiga toxin-producing *Escherichia coli* O157:H7 is a food-borne pathogen and the major cause of hemorrhagic colitis. *Pseudomonas* is the genus most frequent psychrotrophic spoilage microorganisms present in milk. Two-species bacterial systems with *E. coli* O157:H7, non-pathogenic *E. coli*, and *P. fluorescens* in skimmed milk at 7, 13, 19, or 25ºC was studied. Bacterial interactions were modelled after applying a Bayesian approach. No direct correlation between *P. fluorescens*'s growth rate and its effect on the maximum population densities of *E. coli* species was found. The results show the complexity of the interactions between two species into a food model. The use of natural microbiota members to control foodborne pathogens could be useful to improve food safety during the processing and storage of refrigerated foods.

# 1. Introduction

Shiga toxin-producing *Escherichia coli* O157:H7 strains are foodborne pathogens causing hemorrhagic colitis or the hemolytic uremic syndrome (Karmali, 1989). These microorganisms can be transmitted through consumption of undercooked meat, vegetables, contaminated water, unpasteurized dairy products, and raw milk (M. P. Doyle, Zhao, Meng, & Zhao, 1997; Duncan et al., 1986; Griffin & Tauxe, 1991; Louise Martin et al., 1986). The survival capacity of *E. coli* O157:H7 can goes as far as several days or weeks in milk and dairy products (Arocha, McVey, Loder, Rupnow, & Bullerman, 1992; Dineen, Takeuchi, Soudah, & Boor, 1998; Hudson, Chen, Hill, & Griffiths, 1997) showing the importance of post processing contamination and the associated health risks.

Soil, water and vegetation are the main sources of psychrotrophic spoilage microorganisms, such as *Pseudomonas*, to milk (Cousin, 1982; De Jonghe et al., 2011; de Oliveira, Favarin, Luchese, & McIntosh, 2015). During the pre-processing period (3-4 days) prior to pasteurization, psychrotrophic bacteria can grow and cause significant chemical changes (De Jonghe et al., 2011; Farrag & Marth, 1989a). *Pseudomonas* also appears in pasteurized dairy products as a post-processing contaminant (Chiesa et al., 2014; Cousin, 1982).

Different bacterial species interact without physical barriers in many natural environments, and foods are one such example where co-culture experiments have shown the prevailing genotypes (Avendaño-Pérez & Pin, 2013; Cornforth & Foster, 2013; Dubey & Ben-Yehuda, 2011; Nadell & Bassler, 2011). Spoilage microorganisms are able to enhance, limit or be neutral on the growth of pathogenic species (Buchanan & Bagi, 1999). *P. fluorescens* produces extracellular materials resulting in a competitive advantage over other species (Nadell & Bassler, 2011) such that the competitor is physically displaced (Schluter, Nadell, Bassler, & Foster, 2015) or the nutrients access impeded (Kim, Racimo, Schluter, Levy, & Foster, 2014). In early research on the topic, Graves & Frazier (1963) and Seminiano & Frazier (1966) reported the enhancement of *Staphylococcus aureus*'s growth in the presence of *Pseudomonas* spp. Marshall & Schmidt (1991; 1988) and (Farrag & Marth (1989a) found similar results when *Listeria monocytogenes* was co-cultured in the presence of *P. fluorescens*. Other studies reported conversely that *P. fluorescens* can inhibit the growth of *L. monocytogenes* (Al-Zeyara, Jarvis, & Mackey, 2011; Buchanan & Bagi, 1999; Buchanan & Bagi, 1997; Cheng, Doyle, & Luchansky, 1995; Farrag & Marth, 1989b; Fgaier & Eberl, 2010; Freedman, Kondo, & Willrett, 1989; Mellefont, McMeekin, & Ross, 2008). The specific addition of glucose to TSB broth stimulated the inhibition of *E. coli* O157:H7 by *P. fluorescens* (Samelis & Sofos, 2002). Liao (2009) and (Liao, Cooke, & Niemira (2010) reported that *P. fluorescens* and *Bacillus* spp. were able to act as biocontrol agents of *Salmonella* Saintpaul on Jalapeno pepper or of *E. coli* O157:H7 on TSA agar and bell pepper disks. On the surface of spinach leaves, Olanya, Annous, Niemira, Ukuku, & Sommers (2012) reported that *P. fluorescens* moderately suppressed the growth of *E. coli* O157:H7. The same research group also showed *P. fluorescens* inhibition of *E. coli* O157:H7 in poor environments such as distilled water or buffered peptone water (Olanya, Ukuku, & Niemira, 2014).

    The aim of our work is to study the interaction between co-culturing bacterial species using Bayesian inference. The Bayesian approach provides a consistent framework for estimating parameters from a model using prior knowledge about the system to improve the estimations (Rickett et al., 2015), with the advantage of including uncertainties within the model (Chatzilena, van Leeuwen, Ratmann, Baguelin, & Demiris, 2019). Two-species systems were co-cultured with *E. coli* O157:H7 or non-

pathogenic *E. coli* and *P. fluorescens* in skimmed milk at several temperatures.

**2. Material and methods**

*2.1. Bacterial cultures and inoculation*

Three strains of *Escherichia coli*: O157:H7 LCDC 86-51 (EcO1; Shiga toxin-producing strain isolated from hemorrhagic colitis, Canada), O157:H7 ATCC 35150 (EcO2; Shiga toxin-producing strain isolated from hemorrhagic colitis, USA), and the non-pathogenic strain *E. coli* ATCC 8739 (Ec) were used. All strains were cultured overnight in BHI broth (Difco, BD Diagnostics, Franklin Lakes, NJ, USA) at 37 ºC. *Pseudomonas fluorescens* ATCC 13525 (Pf; isolated from pre-filter tanks and town water works, UK) was cultured in BHI at 25 ºC for 24 h.

Bacterial cultures were grown at 37 ºC until a population of $10^9$ CFU/ml was reached as previously described (Quinto, Marín, Caro, Mateo, & Schaffner, 2018). Briefly, serial dilutions were prepared in 0.1% sterile peptone water (Difco) and 1 ml from adequate dilutions were added to 250 ml screw-capped Erlenmeyer flasks containing 100 ml of 10% reconstituted sterile skimmed milk. Populations of ca. $10^4$ CFU/ml were achieved, and evaluated by spreading onto TSA (Difco) plates and incubating at 37°C for 48 h.

*2.2. Co-cultures and enumeration*

A first 4-level factor (EcO1, EcO2, Ec, and Pf strains) and a second 4-level factor (7, 13, 19 or 25 ºC) were used for a 4 x 4 full factorial experiment. Starting concentrations of ca. $10^4$ CFU/ml were selected and co-cultures of EcO1, EcO2, Ec, and Pf were prepared and stored at 7, 13, 19 or 25 °C. Actual co-culture starting populations of 3.9-4.1, 4.0-4.1, 4.0-4.1, and 3.9-4.1 log CFU/ml were obtained for EcO1, EcO2, Ec, and Pf, respectively. Single cultures of EcO1, EcO2, Ec, and Pf were cultivated with the same initial populations. Control cultures of un-inoculated skimmed milk were prepared and stored under the same conditions.

Cultures were sampled exceeding the shelf life of pasteurized milk at 0, 2, 4, 6, 12, or 24 h, and 2, 4, 8, 12, 16, 20, 28 days. Aliquots (0.1 ml) were surface-plated onto MacConkey Sorbitol Agar (Difco) and Fluorocult VRB-Agar (Merk, Darmstadt, Germany) or onto Pseudomonas Agar F or Flo Agar (Difco). For *Escherichia* spp. counting, plates were incubated at 37 °C for 18-24 h, and random colonies were serologically confirmed using the *E. coli* O157 Latex Test Kit (Oxoid, Thermo Fisher Scientific, UK). *P. fluorescens* colonies were counted after an incubation at 35 ºC for 24-48 h.

*2.3. Bayesian modeling of microbial interactions*

A Bayesian estimation of the parameters was considered. A modified generic primary growth model (M. Cornu, Billoir, Bergis, Beaufort, & Zuliani, 2011) was selected:

$$\frac{dN_t/dt}{N_t} = \frac{d(ln(N_t))}{dt} = \mu_{max}\alpha_t f_t \tag{1}$$

where $(dN_t/dt)/N_t$ is the relative or instantaneous growth rate of the microorganism, $N_t$ is the cell concentration in a bacterial culture at time t, and $\mu_{max}$ is the maximum growth rate. The term $\alpha_t$ is an adjustment function, and $f_t$ is a logistic inhibition function for two-species mixed cultures (M. Cornu, 2001):

$$\alpha_t = \begin{cases} 0, & t < \lambda \\ 1, & t \geq \lambda \end{cases}$$

$$f_t = 1 - \frac{Na_t + Nb_t}{N_{max}} \tag{2}$$

where $\lambda$ is the lag time, $Na_t$ and $Nb_t$ are the cell concentration of the microorganisms a or b in co-culture at time t, and $N_{max}$ is the total carrying capacity (both species). For EcO1 cultures the model can be re-defined:

$$\frac{dEcO1/dt}{EcO1_t} = \frac{d(\ln(EcO1_t))}{dt} = \mu_{EcO1}\alpha_t\left(1 - \frac{EcO1_t}{EcO1_{max}}\right) \tag{3a}$$

$$\frac{dEcO1/dt}{EcO1_t} = \frac{d(\ln(EcO1_t))}{dt} = \mu_{EcO1(Pf)}\alpha_t\left(1 - \frac{EcO1_t + Pf_t}{N_{max}}\right) \tag{3b}$$

where $\mu_{EcO1}$ (3a) and $\mu_{EcO1(Pf)}$ (3b) are the maximal growth rates of EcO1 cultured alone or in the presence of *P. fluorescens*, respectively. Similar approaches to the equations (3a-b) were done for the cultures of EcO2 ($\mu_{EcO2}$, $\mu_{EcO2(Pf)}$), Ec ($\mu_{Ec}$, $\mu_{Ec(Pf)}$), and *P. fluorescens* ($\mu_{Pf}$, $\mu_{Pf(EcO1)}$, $\mu_{Pf(EcO2)}$, and $\mu_{Pf(Ec)}$). When the cultures reached their maximal values, a decline period was observed. The decline phase was modeled alone with a modification of equations (3a-b), i.e., $\mu$ was replaced by the negative-sign parameter k in order to reflect the negative slope of that survival growth section.

The approach above assumes deterministic behavior, but an error term may be introduced to reflect the influence of factors outside the experimental design. Thus, the observed concentration of bacteria at time t may modelled as $N_t^* = N_t + \varepsilon_t$, where $N_t$ is the population of EcO1, EcO2, Ec, or Pf cultured alone or in co-culture, and $\varepsilon_t$ is a normally distributed error term with zero mean and constant variance equal to $\sigma_t$: $N_t^* \sim$ Normal ($N_t$, $\sigma_t$).

A Bayesian estimation of the parameters for computing the posterior distribution of parameters of the model was carried out. The estimated parameters are shown in Figure 1 as circles: the growth rates of the microorganisms cultured alone ($\mu_{EcO1}$, $\mu_{EcO2}$, $\mu_{Ec}$, and $\mu_{Pf}$), the 2-species mixtures ($\mu_{EcO1(Pf)}$, $\mu_{EcO2(Pf)}$, $\mu_{Ec(Pf)}$, $\mu_{Pf(EcO1)}$, $\mu_{Pf(EcO2)}$, and $\mu_{Pf(Ec)}$), and the standard deviation of errors ($\sigma_t$); the other terms are constants and are shown as squares. Decline rates and the 95% credible intervals were also estimated based on the posterior distribution of parameters from equations 3a-b.

The Runge-Kutta method was used to discretize the system of differential equations (Dormand & Prince, 1980); then the system was included in a probabilistic model and the Hamiltonian Monte Carlo method (HMCM) was used for parameters estimation (Vinet & Zhedanov, 2010) generating samples from the posterior distributions of parameters $\mu_t$ and $\sigma_t$ (Carpenter et al., 2017). R (R-Project, 2014) via Rstan (Stan Development Team, 2017) was used for algorithmic programming. Codes are available from author JMM. Orange 3.20.1 for macOs freely available at https://orange.biolab.si (Demšar et al., 1967) was used for displaying data in a mosaic plot.

*2.4. Estimation of the maximum population density and the time to reach a population*

Plate counts of *E. coli* strains and *P. fluorescens* were transformed to decimal logarithmic values. The maximum population density ($N_{max}$) was estimated for each culture using the DMFit Web Edition, ComBase (Baranyi & Roberts, 1994). The estimated growth curves were used to obtain the time to reach (ttr) populations of 6 or 8 log CFU/ml.

## 3. Results and discussion

*3.1. Bayesian modelling of microbial interactions*

Representative examples of the Bayesian inference of growth and decline periods for *E. coli* O157:H7 LCDC 86-51 (EcO1), *E. coli* O157:H7 ATCC 35150 (EcO2), and non-pathogenic *E. coli* (Ec) co-cultured with *P. fluorescens* (Pf) at 7 or 25 ºC in skimmed milk are shown in Figures 2, S1, S3, and Figures 3, S2, S4, respectively. Table 1 shows posterior means of the parameters and the limits of the credible intervals (2.5% and 97.5%) of growth rates ($\mu_{EcO1}$, $\mu_{EcO2}$, $\mu_{Ec}$, or $\mu_{Pf}$) and decline rates ($k_{EcO1}$, $k_{EcO2}$, $k_{Ec}$, or $k_{Pf}$) of *E. coli* spp. and *P. fluorescens* cultured alone or in co-cultures. The *E. coli* spp. co-cultures show the lowest µ values at 7 and 13 ºC ranging from 0.563 d$^{-1}$ for the EcO1(Pf) co-cultures at 7 ºC to 0.945 d$^{-1}$ for the Ec(Pf) co-cultures at 13 ºC. At 19 and 25 ºC the µ values were similar ranging from 1.358 d$^{-1}$ for the EcO1(Pf) co-cultures to 2.178 d$^{-1}$ for the Ec(Pf) co-cultures at 25 ºC. Figure 4A shows the µ estimates using colour codes. The highest µ values were detected at 19 and 25 ºC (green and orange colours, respectively). It is interesting to observe the differences on the growth rates between *E. coli* spp. and *P. fluorescens* cultured alone: a psychrotrophic bacteria such as *P. fluorescens* did not show blue colour (the lowest values) at any temperature; however, *E. coli* spp. showed it at 7 and 13 ºC. Co-cultured *P. fluorescens* showed similar code colours as when cultured alone, never showing the blue code. Figure 4B shows the µ estimates of *E. coli* spp. co-cultured with *P. fluorescens*, showing that the effect of *P. fluorescens* on the growth rate of *E. coli* strains appears to be greater at low temperatures (7 and 13 ºC) increasing *E. coli* spp. µ values. At higher temperatures (19 and 25 ºC) *P. fluorescens* does not seems to cause the same effect.

At 7 ºC decreasing populations of the three *E. coli* strains cultured alone were not detected (Table 1). Decreasing populations (k values) were not found for EcO2 and Ec strains single-cultured, and for EcO1(Pf), EcO2(Pf) and Ec(Pf) co-cultures at 13 ºC. The positive k values are included within the Highest Posterior Density (HPD) intervals between a negative 2.5% interval value and a positive 97.5% interval value. As the zero value is into the interval, the estimated values are not significantly different from zero, i.e., there is not growth nor decline with a 97.5% of confidence, and the populations are stable. The fastest decline rates were observed in the EcO1(Pf) and Ec(Pf) co-cultures at 19 ºC ($-1.804$ $d^{-1}$ and $-1.709$ $d^{-1}$, respectively). The higher decline rates from single-cultured strains were found in EcO2 cultures at 19 ºC ($-0.233$ $d^{-1}$) and 25 ºC ($-0.247$ $d^{-1}$).

Table 2 shows posterior means of the parameters and the limits of the credible intervals (2.5% and 97.5%) of the standard deviations for growth rates ($\sigma_{EcO1}$, $\sigma_{EcO2}$, $\sigma_{Ec}$, or $\sigma_{Pf}$) and decline rates ($-\sigma_{EcO1}$, $-\sigma_{EcO2}$, $-\sigma_{Ec}$, or $-\sigma_{Pf}$) of *E. coli* spp. and *P. fluorescens* cultured alone or in co-cultures. Standard deviation is read as the predicted concentrations' random error of the microorganisms' real observations.

*3.2. Estimation of the $N_{max}$ and the ttr*

The maximum population density ($N_{max}$) of *E. coli* spp. and *P. fluorescens* in single cultures or co-cultured in milk are shown in Table 3. The lowest *E. coli* spp. $N_{max}$ values were observed at 7ºC (4.4–4.8 log CFU/ml in single cultures and 5.2–5.4 log CFU/ml in co-cultures); at 13, 19, or 25 ºC, the $N_{max}$ values were similar for all co-cultures (8.0–8.5 log CFU/ml in single cultures and 7.9–8.5 log CFU/ml in co-cultures). The *P. fluorescens* $N_{max}$ values were similar for all co-cultures at all temperatures (8-9 log CFU/mL).

The time to reach (ttr) a population of 6 or 8 log CFU/ml is shown in Table 3. These populations were found just before the carrying capacities were reached, and both fall within the linear period of the exponential growth where rates show a Log-Normal distribution regardless of the environmental conditions and the initial population of microorganisms (Akkermans, Logist, & Van Impe, 2018; METRIS, 2003; Pin & Baranyi, 2006). At 7ºC *E. coli* spp. did not reach 6 log CFU/ml whether single-cultured or co-cultured. At 13 ºC all *E. coli* spp. cultures reached 6 log CFU/ml at 1.60–2.24 d; similar results were found at 19 or 25 ºC with lower ttr 6 log values showing a faster growth: 0.56–0.64 d or 0.32–0.40 d, respectively. All *P. fluorescens* cultures reached ca. 6 log CFU/ml after 0.56–0.72 d at 7 ºC; the ttr 6 log results at 13 ºC were slightly higher (0.64-1.12 d), decreasing at 19 and 25 ºC and showing faster growth: 0.56–0.64 d and 0.40–0.48 d, respectively. The ttr 6 log values of *P. fluorescens* were lower than those from *E. coli* strains at all temperatures, indicating faster growth. All *E. coli* spp. and *P. fluorescens* cultures were able to reach a population of 8 log CFU/ml, except *E. coli* spp. at 7 ºC. At 13 ºC all *E. coli* spp. cultures reached 8 log CFU/ml at 3.36–3.68 d; lower results were found at 19 or 25 ºC indicating a faster growth: 1.04–1.12 d or 0.56–0.96 d, respectively. All *P. fluorescens* cultures reached at 8 log CFU/ml after 0.96–1.04 d at 7 ºC; the ttr 8 log results at 13, 19, or 25 ºC were lower showing a faster growth: 1.44–1.76 d, 0.96–1.04 d, or 0.56–0.88 d, respectively. The *P. fluorescens*'s ttr 8 log values were lower than those from *E. coli* strains at all temperatures, indicating overall faster growth.

*3.3. Discussion*

The presence of *E. coli* O157:H7 strains in refrigerated food such as milk depicts a health risk for the consumers. The native microbiota or the presence of protective cultures could compete with the pathogens and help in controlling *E. coli* O157:H7 strains during the processing and storage of refrigerated food (Park, Worobo, & Durst, 2001; Samelis & Sofos, 2002). *Pseudomonas* spp. could be important competitors in perishable refrigerated food products due to their psychrotrophic profile (able to grow at 0–15 °C) (Samelis & Sofos, 2002; Ternström, Lindberg, & Molin, 1993).

      Antagonistic microorganisms (e.g. *Pseudomonas* spp.), may be useful in the control of *E. coli* O157:H7 growth. According to Samelis & Sofos (2002), *E. coli* O157:H7 co-cultured with *Pseudomonas* sp. grew faster as the temperature increased from 10 to 15 or to 25°C in TSB broth. These authors found that the pathogen inhibition was enhanced in co-cultures grown at 10 to 15°C with 1% of added glucose. At 25 ºC the inhibition was enhanced even without added glucose. Previously, Janisiewicz, Conway, & Leverentz (1999) reported that *P. syringae* inoculated into apple injuries inhibited the growth of *E. coli* O157:H7. Similar results were found by Vold, Holck, Wasteson, & Nissen (2000) and Nissen, Maugesten, & Lea (2001) when a high level of ground beef native flora inhibited the growth of *E. coli* O157:H7 at 10–12°C. Samelis & Sofos (2002) found that the maximum population density of *E. coli* O157:H7 was suppressed in co-culture with *Pseudomonas* at 10, 15, and 25 °C. These results are in agreement with the Jameson Effect (Jameson, 1962); indeed, the inhibition of a population not in its stationary phase by another population in it is observed (Jameson, 1962; Mellefont et al., 2008). Similar results were found by Buchanan & Bagi (1999) when *P. fluorescens* suppressed *Listeria monocytogenes* growth by inhibiting its maximum population density at low incubation temperatures (4ºC); the inhibition was less evident at higher temperatures (12 and 19 ºC). McKellar (2007) also reported that a raw milk isolate of *P. fluorescens* suppressed the growth of *E. coli* O157:H7 in nutrient broth at 22 ºC only when *P. fluorescens* had reached its maximum population. Similar

results were found in a previous study in co-cultures of *Listeria* spp. with *P. fluorescens* (Quinto et al., 2018) but without differences between low and high temperatures, as Buchanan & Bagi (1999) did. Samelis & Sofos (2002) reported that *E. coli* O157:H7 co-cultured with *Pseudomonas* reached a population of ca. 6 log CFU/ml after ~2.6 d at 10 ºC, not achieving a population of 8 log CFU/ml along the study (14 d); the same co-culture reached a maximum of 7 log CFU/ml (~6.1 d) when the TSB broth was supplemented with 1% of glucose. In our study *E. coli* spp. did not reach 6 or 8 log CFU/ml at 7 ºC. Samelis & Sofos (2002) found that *E. coli* O157:H7 co-cultured at 15 ºC with *Pseudomonas* reached populations of 6 or 8 log CFU/ml after ~0.5 or 2 d, respectively and when the co-cultures were supplemented with 1% of glucose *E. coli* O157:H7 achieved populations of 6 or 8 log CFU/ml after ~1.2 or 7 d, respectively. We found slightly slower growth in our study at 13 ºC with a ttr 6 log of 1.6–2.2 d, and a ttr 8 log of 3.5–3.7 d. These authors (Samelis & Sofos, 2002) did not detect changes in pH along the incubation period (14 d) irrespective of the temperature and the type of culture: pH values of 7.3–7.4; in contrast, pH reductions were pronounced when 1% of glucose was added to the medium decreasing to values of 5.0–6.0. The pH values in single cultures or in co-cultures in our study decreased along the study from 6.7–6.8 to about 6.5 after 28 d at 7 or 13 ºC (data not shown). At 19 or 25 ºC the pH decreased until values of about 4.0–4.5 (data not shown) at the end of the study probably due to *E. coli* use of the lactose from the milk.

Lebert, Robles-Olvera, & Lebert (2000) found that the growth of *L. monocytogenes* and *L. innocua* were not affected by *Pseudomonas* spp. at 6 °C on decontaminated meat but *Pseudomonas* spp. did affect *L. innocua* on native-microbiota contaminated meat – when *Pseudomonas* achieved their stationary phase *Listeria* was able to grow. These results are in contrast with previous (Cornu, Kalmokoff, &

Flandrois, 2002; Besse et al., 2010; Quinto et al., 2018) and current results which found that *P. fluorescens* exerts similar inhibitory effects on the *E. coli* strains studied. Besse et al. (2010) noted interactions at the end of the exponential phase – when a strain reached its carrying capacity the growth of both strains stopped. McKellar (2007) reported that nutrient limitation was the cause of the competition between *Pseudomonas* and *E. coli* O157:H7. But quorum sensing stimuli has also been suggested as a mechanism (Chu et al., 2012; Diggle, Griffin, Campbell, & West, 2007; Dubey & Ben-Yehuda, 2011; Ng & Bassler, 2009; West, Griffin, Gardner, & Diggle, 2006). Once a faster growing microorganism reaches its maximum population, the production of signaling molecules also reaches its maximum, indicating to other species of the mixed culture that the carrying capacity of the culture has been achieved. Chu et al. (2012) also showed how *E. coli* indole production inhibited *P. aeruginosa* factors important for competition.

As *P. fluorescens* constitutes a major component of native bacteria associated with fresh and minimally processed produce, Liao (2009) studied the control of foodborne pathogens by *P. fluorescens* AG3A and *Bacillus* YD1 both isolated from fresh peeled baby carrots. Both strains reduced the growth of *L. monocytogenes, Yersinia enterocolitica, Salmonella enterica*, and *E. coli* O157:H7 at 20 ºC but not at 10 ºC. Olanya et al. (2012) reported a moderate inhibition of *E. coli* O157:H7 by *P. fluorescens* on spinach leaves surfaces. These strains showed similar behaviors when they were co-cultured with nutrient restrictions at 10-35 ºC for 48 h (Olanya et al., 2014); these authors found an *E. coli* O157:H7 ttr 6 log of 1.5 d at 20 ºC, without reaching a population of 8 log CFU/ml; at 35 ºC the ttr 6 log was about 1.1 d, and the ttr 8 log 1.6 d.

In our experiments *P. fluorescens* grew faster than *E. coli* spp. at 7 and 13 °C cultured alone as well as in co-culture, with higher $\mu_{Pf}$ values and lower ttr 6 or 8 log; however, *P. fluorescens* did not affect the $\mu_{EcO1}$, $\mu_{EcO2}$, and $\mu_{Ec}$ values. Similar behavior was observed when *L. monocytogenes* were co-cultured with *Lactobacillus sakei* (Quinto, Marín, & Schaffner, 2016) or *P. fluorescens* (Quinto et al., 2018) together with higher $N_{max}$ of both competitors; in contrast the current study did not find higher $N_{max}$ in *P. fluorescens* single cultured or co-cultured with *E. coli* spp. At 19 and 25 °C the ttr 6 or 8 log of *P. fluorescens* were lower than those of *E. coli* spp. and the $\mu_{Pf}$ values were similar between single cultures and co-cultures showing also similar $N_{max}$: slight maximum population increases (< 1 log CFU/ml) of *P. fluorescens* at 25 °C were observed. These results are not consistent with the Jameson Effect (Jameson, 1962; Mellefont et al., 2008; Ross, 2000) with regard to the inhibition of one species by another that has reached the stationary phase; indeed, the Jameson-effect hypothesis states that the competition in food mixed populations is restricted to the limitation of the maximum population, with no effect on the lag time or the growth rate. Our work supports the Jameson-effect hypothesis as neither lag times nor growth rates of *E. coli* species seems to be affected by *P. fluorescens*. However, the maximum population densities we observed do not support the Jameson-effect hypothesis. There is no correlation between the $\mu_{Pf}$ and its effect on the maximal population densities of *E. coli* spp. (Pearson's coefficient correlation of –0.407). The values of *E. coli* spp. $N_{max}$ were high at all temperatures except at 7 °C ($N_{max}$ of about 5.2–5.4 log CFU/ml), so the increasing $\mu_{Pf}$ values did not increased *E. coli* spp. $N_{max}$ together with the increase of the temperatures. It would be possible to consider the fermentation of milk lactose by the *E. coli* spp. as a "high risk, high reward" strategy in the two-species communities studied (Mao, Blanchard, & Lu, 2015; Stubbendieck, Vargas-Bautista, & Straight, 2016). *E.*

*coli* spp. must engage additional competitive mechanisms to remain viable such as lactose fermentation (Liu et al., 2011) although it was far from the aim of this work to explore it. These interactions could also be related to physical location or resource usage overlapping between both populations (Stubbendieck et al., 2016). Another possible explanation for the absence of the Jameson effect at the higher temperatures studied could be a "counterattack strategy". Some authors have reported that *P. aeruginosa* suffering the attack from *Vibrio cholerae* or *Acinetobacter baylyi*'s type VI secretion system (T6SS) respond striking back with its own T6SS (Marek Basler, Ho, & Mekalanos, 2013). The T6SS is a multiprotein contractile-weapon complex that participates in interbacterial competition delivering toxins into both prokaryotic and eukaryotic cells. The T6SS complex does occur in *Escherichia coli* and *Salmonella* (Journet & Cascales, 2016) including enterohemorrhagic *E. coli* O157:H7 (Wan et al., 2017). Decoin et al. (2014) described a T6SS involved in *P. fluorescens* bacterial competition against the potato tuber pathogen *Pectobacterium atrosepticum*. Although the objective of our study is far from the description of a T6SS *P. fluorescens* activity against *E. coli* spp., the results provide evidence for a bacterial ''tit-for-tat'' (Sachs, Mueller, Wilcox, & Bull, 2004) or "T6SS dueling" (Basler & Mekalanos, 2012) evolutionary strategies that control interactions among different bacterial species.

*3.4. Conclusions*

The aim of this work was to study and model the dynamics of the competition between *Escherichia coli* O157:H7 and *Pseudomonas fluorescens* co-cultured at 7, 13, 19, and 25 °C in milk. A parametric Bayesian approach was used assuming that the parameters μ (growth rate), k (decline rate), σ (standard deviation of the growth rates), and –σ (standard deviation of the decline rates) are random variables with their own prior distributions. Model results and confidence intervals are based on a probabilistic background. The highest *E. coli* O157:H7 populations were similar at all temperatures, except at 7 ºC: *E. coli* spp. strains reached their maximal population of 4 log CFU/ml cultured alone, and 5 log CFU/ml co-cultured with *P. fluorescens*. At 13, 19, and 25 ºC *E. coli* spp. reached their maximal population of 8 log CFU/ml single cultured and co-cultured, with times to reach a population of 6 log CFU/ml after ~48 h at 13 ºC or ~24 h at 19 and 25 ºC. *P. fluorescens* achieved its maximal densities of 8–9 log CFU/ml in all cultures at all temperatures, with similar times to reach a population of 6 or 8 log CFU/ml. The results obtained show that the growth rate of *P. fluorescens* has no direct correlation with its effect on the maximal population of *E. coli* strains. Modeling the behavior of bacterial communities helps in understanding their dynamics. The inhibition of foodborne pathogens with the use of some species from the natural food microbiota as probiotics may be a tool to improve the safety of refrigerated foods such as milk and dairy products.

**Acknowledgements**

This research did not receive any specific grant from funding agencies in the public, commercial, or not-for-profit sectors.

for microorganisms. *Nature Reviews Microbiology*, *4*(8), 597–607.

http://doi.org/10.1038/nrmicro1461

Table 1. Bayesian estimates of the posterior means and Highest Posterior Density intervals (HPD: 2.5 and 97.5%) of the growth ($\mu$; h$^{-1}$) and decline (k; h$^{-1}$) rates of *E. coli* O157:H7 LCDC 86-51, *E. coli* O157:H7 ATCC 35150, non-pathogenic *E. coli*, and *P. fluorescens* cultured alone (EcO1, EcO2, Ec, Pf), or co-cultured (EcO1 + Pf, EcO2 + Pf, Ec + Pf) at 7, 13, 19 or 25ºC.

| Cultures | Temp (ºC) | Strain | μ mean | HPD intervals 2.5% | HPD intervals 97.5% | k mean | HPD intervals 2.5% | HPD intervals 97.5% |
|---|---|---|---|---|---|---|---|---|
| EcO1, EcO2, Ec, Pf | 7 | EcO1 | 0.104 | 0.041 | 0.261 | – | – | – |
| | | EcO2 | 0.123 | 0.014 | 0.789 | – | – | – |
| | | Ec | 0.120 | – 0.019 | 1.382 | – | – | – |
| | | Pf | 1.690 | 1.173 | 2.202 | 0.485 | – 0.514 | 2.160 |
| | 13 | EcO1 | 0.535 | 0.356 | 0.782 | – 0.017 | – 2.003 | 1.987 |
| | | EcO2 | 0.503 | 0.348 | 0.727 | 0.520 | – 0.438 | 2.148 |
| | | Ec | 0.494 | 0.345 | 0.688 | 0.007 | – 1.937 | 1.965 |
| | | Pf | 0.990 | 0.633 | 1.450 | 0.460 | – 0.488 | 2.007 |
| | 19 | EcO1 | 1.664 | 1.264 | 2.070 | – 0.164 | – 0.190 | – 0.142 |
| | | EcO2 | 1.597 | 1.006 | 2.214 | – 0.233 | – 0.268 | – 0.203 |
| | | Ec | 1.577 | 0.919 | 2.232 | – 0.212 | – 0.240 | – 0.187 |
| | | Pf | 1.539 | 0.981 | 2.126 | – 0.056 | – 0.072 | – 0.045 |
| | 25 | EcO1 | 2.550 | 1.505 | 3.336 | – 0.179 | – 0.209 | – 0.154 |
| | | EcO2 | 2.781 | 1.591 | 3.609 | – 0.247 | – 0.285 | – 0.215 |
| | | Ec | 2.007 | 1.714 | 2.275 | – 0.170 | – 0.194 | – 0.150 |
| | | Pf | 1.775 | 1.185 | 2.348 | – 0.128 | – 0.147 | – 0.111 |
| EcO1 + Pf | 7 | EcO1 | 0.563 | 0.235 | 0.918 | – 0.281 | – 0.447 | 0.538 |
| | | Pf | 2.021 | 1.357 | 2.749 | – 0.036 | – 0.116 | 0.361 |
| | 13 | EcO1 | 0.692 | 0.425 | 1.026 | 0.242 | – 1.292 | 2.028 |
| | | Pf | 0.944 | 0.670 | 1.218 | 0.256 | – 1.247 | 2.005 |
| | 19 | EcO1 | 1.596 | 1.138 | 2.020 | – 1.804 | – 2.467 | – 1.264 |
| | | Pf | 1.579 | 1.049 | 2.025 | – 0.064 | – 0.140 | – 0.025 |
| | 25 | EcO1 | 1.358 | – 0.900 | 3.259 | – 0.262 | – 0.304 | – 0.225 |
| | | Pf | 1.167 | – 0.869 | 2.628 | – 0.131 | – 0.157 | – 0.108 |
| EcO2 + Pf | 7 | EcO2 | 0.824 | 0.357 | 1.208 | – 0.017 | – 1.841 | 2.009 |
| | | Pf | 2.006 | 1.406 | 2.645 | 0.202 | – 1.108 | 1.952 |
| | 13 | EcO2 | 0.664 | 0.420 | 0.952 | 0.130 | – 1.293 | 2.017 |
| | | Pf | 0.936 | 0.660 | 1.205 | 0.163 | – 1.332 | 1.989 |
| | 19 | EcO2 | 1.624 | 1.226 | 2.019 | – 0.260 | – 0.300 | – 0.224 |
| | | Pf | 1.641 | 1.239 | 2.047 | – 0.046 | – 0.062 | – 0.030 |
| | 25 | EcO2 | 1.857 | 0.187 | 2.791 | – 1.371 | – 2.210 | – 0.304 |
| | | Pf | 1.782 | 0.306 | 2.738 | – 0.698 | – 2.129 | – 0.168 |
| Ec + Pf | 7 | Ec | 0.581 | 0.357 | 0.791 | – 0.222 | – 1.738 | 1.835 |
| | | Pf | 1.679 | 0.855 | 2.227 | – 0.019 | – 1.174 | 1.856 |
| | 13 | Ec | 0.945 | 0.661 | 1.270 | 0.266 | – 1.468 | 2.104 |
| | | Pf | 1.204 | 0.854 | 1.487 | 0.319 | – 1.305 | 2.078 |
| | 19 | Ec | 1.495 | 1.114 | 1.907 | – 1.709 | – 2.064 | – 1.089 |
| | | Pf | 1.606 | 1.185 | 2.030 | – 0.072 | – 0.156 | – 0.038 |
| | 25 | Ec | 2.178 | 1.667 | 2.620 | – 0.143 | – 0.311 | – 0.065 |
| | | Pf | 2.575 | 1.957 | 3.149 | – 0.095 | – 0.128 | – 0.066 |

Table 2. Bayesian estimates of the posterior means and Highest Posterior Density intervals (HPD: 2.5 and 97.5%) of the standard deviations for growth (σ) and decline (−σ) periods of *E. coli* O157:H7 LCDC 86-51, *E. coli* O157:H7 ATCC 35150, non-pathogenic *E. coli*, and *P. fluorescens* cultured alone (EcO1, EcO2, Ec, Pf), or co-cultured (EcO1 + Pf, EcO2 + Pf, Ec + Pf) at 7, 13, 19 or 25°C.

| Cultures | Temp (°C) | Strain | σ mean | HPD intervals 2.5% | HPD intervals 97.5% | − σ mean | HPD intervals 2.5% | HPD intervals 97.5% |
|---|---|---|---|---|---|---|---|---|
| EcO1, EcO2, Ec, Pf | 7 | EcO1 | 0.234 | 0.151 | 0.376 | – | – | – |
| | | EcO2 | 0.292 | 0.182 | 0.499 | – | – | – |
| | | Ec | 0.388 | 0.240 | 0.666 | – | – | – |
| | | Pf | 0.554 | 0.291 | 1.111 | 0.749 | 0.397 | 1.492 |
| | 13 | EcO1 | 0.605 | 0.350 | 1.090 | 0.990 | 0.452 | 2.212 |
| | | EcO2 | 0.552 | 0.322 | 1.000 | 0.891 | 0.406 | 1.986 |
| | | Ec | 0.528 | 0.308 | 0.932 | 1.107 | 0.469 | 2.239 |
| | | Pf | 0.662 | 0.378 | 1.219 | 0.275 | 0.125 | 0.661 |
| | 19 | EcO1 | 0.398 | 0.206 | 0.826 | 0.664 | 0.317 | 1.466 |
| | | EcO2 | 0.619 | 0.329 | 1.228 | 0.724 | 0.360 | 1.565 |
| | | Ec | 0.658 | 0.346 | 1.302 | 0.621 | 0.302 | 1.352 |
| | | Pf | 0.598 | 0.312 | 1.199 | 0.465 | 0.211 | 1.104 |
| | 25 | EcO1 | 0.578 | 0.275 | 1.302 | 0.684 | 0.328 | 1.497 |
| | | EcO2 | 0.536 | 0.241 | 1.290 | 0.728 | 0.352 | 1.575 |
| | | Ec | 0.210 | 0.104 | 0.460 | 0.563 | 0.271 | 1.258 |
| | | Pf | 0.641 | 0.333 | 1.298 | 0.600 | 0.286 | 1.342 |
| EcO1 + Pf | 7 | EcO1 | 0.447 | 0.247 | 0.860 | 0.485 | 0.173 | 1.493 |
| | | Pf | 0.729 | 0.409 | 1.346 | 0.281 | 0.098 | 0.806 |
| | 13 | EcO1 | 0.862 | 0.483 | 1.575 | 0.765 | 0.380 | 1.586 |
| | | Pf | 0.492 | 0.260 | 0.979 | 0.283 | 0.130 | 0.672 |
| | 19 | EcO1 | 0.581 | 0.291 | 1.181 | 1.930 | 1.056 | 3.555 |
| | | Pf | 0.612 | 0.308 | 1.271 | 0.707 | 0.298 | 1.626 |
| | 25 | EcO1 | 1.263 | 0.401 | 2.827 | 0.552 | 0.307 | 1.057 |
| | | Pf | 0.935 | 0.354 | 2.165 | 0.542 | 0.301 | 1.027 |
| EcO2 + Pf | 7 | EcO2 | 0.387 | 0.188 | 0.852 | 1.981 | 1.062 | 3.511 |
| | | Pf | 0.646 | 0.340 | 1.288 | 1.296 | 0.674 | 2.395 |
| | 13 | EcO2 | 0.747 | 0.415 | 1.367 | 0.770 | 0.274 | 1.817 |
| | | Pf | 0.504 | 0.270 | 1.018 | 0.214 | 0.090 | 0.561 |
| | 19 | EcO2 | 0.554 | 0.312 | 1.022 | 0.470 | 0.190 | 1.245 |
| | | Pf | 0.553 | 0.311 | 1.029 | 0.414 | 0.160 | 1.107 |
| | 25 | EcO2 | 0.986 | 0.378 | 2.530 | 1.419 | 0.736 | 2.612 |
| | | Pf | 1.125 | 0.445 | 2.558 | 1.868 | 0.984 | 3.427 |
| Ec + Pf | 7 | Ec | 0.217 | 0.091 | 0.508 | 1.228 | 0.714 | 2.470 |
| | | Pf | 0.645 | 0.297 | 1.524 | 1.695 | 0.816 | 3.205 |
| | 13 | Ec | 0.577 | 0.303 | 1.086 | 0.570 | 0.230 | 1.450 |
| | | Pf | 0.372 | 0.174 | 0.825 | 0.130 | 0.044 | 0.407 |
| | 19 | Ec | 0.588 | 0.333 | 1.103 | 1.764 | 0.915 | 3.434 |
| | | Pf | 0.562 | 0.312 | 1.061 | 0.449 | 0.130 | 1.335 |
| | 25 | Ec | 0.415 | 0.213 | 0.819 | 1.750 | 0.885 | 3.549 |
| | | Pf | 0.760 | 0.443 | 1.354 | 0.653 | 0.270 | 1.637 |

Table 3. Maximal population density ($N_{max}$; log CFU/ml) and time to reach (ttr; d) a population density of 6 or 8 log CFU/ml of *E. coli* spp. strains and *P. fluorescens* cultured alone or in co-culture at 7, 13, 19, or 25ºC.

| Temp (ºC) | Cultures | $N_{max}$ | Time to reach (ttr) a population of | |
|---|---|---|---|---|
| | | | 6 log CFU/ml | 8 log CFU/ml |
| 7 | EcO1 | 4.41 | – | – |
| | EcO2 | 4.42 | – | – |
| | Ec | 4.80 | – | – |
| | Pf | 8.76 | 0.72 | 0.96 |
| | EcO1(Pf) | 5.30 | – | – |
| | EcO2(Pf) | 5.37 | – | – |
| | Ec(Pf) | 5.20 | – | – |
| | Pf(EcO1) | 8.73 | 0.56 | 1.04 |
| | Pf(EcO2) | 8.79 | 0.56 | 1.04 |
| | Pf(Ec) | 9.01 | 0.56 | 1.12 |
| 13 | EcO1 | 8.46 | 2.08 | 3.36 |
| | EcO2 | 8.33 | 2.08 | 3.52 |
| | Ec | 8.25 | 2.24 | 3.68 |
| | Pf | 8.28 | 1.12 | 1.76 |
| | EcO1(Pf) | 8.29 | 2.24 | 3.52 |
| | EcO2(Pf) | 8.29 | 2.08 | 3.52 |
| | Ec(Pf) | 8.20 | 1.60 | 3.68 |
| | Pf(EcO1) | 8.33 | 0.96 | 1.76 |
| | Pf(EcO2) | 8.30 | 0.96 | 1.76 |
| | Pf(Ec) | 8.21 | 0.64 | 1.44 |
| 19 | EcO1 | 8.37 | 0.56 | 1.04 |
| | EcO2 | 8.17 | 0.64 | 1.04 |
| | Ec | 8.04 | 0.64 | 1.04 |
| | Pf | 8.11 | 0.64 | 1.04 |
| | EcO1(Pf) | 8.26 | 0.56 | 1.04 |
| | EcO2(Pf) | 8.23 | 0.56 | 1.04 |
| | Ec(Pf) | 7.89 | 0.64 | 1.12 |
| | Pf(EcO1) | 8.14 | 0.56 | 0.96 |
| | Pf(EcO2) | 8.11 | 0.56 | 1.04 |
| | Pf(Ec) | 8.13 | 0.56 | 1.04 |
| 25 | EcO1 | 8.22 | 0.32 | 0.64 |
| | EcO2 | 8.33 | 0.32 | 0.64 |
| | Ec | 8.28 | 0.40 | 0.96 |
| | Pf | 8.94 | 0.48 | 0.88 |
| | EcO1(Pf) | 8.43 | 0.32 | 0.56 |
| | EcO2(Pf) | 8.54 | 0.40 | 0.56 |
| | Ec(Pf) | 8.10 | 0.36 | 0.80 |
| | Pf(EcO1) | 8.71 | 0.48 | 0.72 |
| | Pf(EcO2) | 8.98 | 0.40 | 0.64 |
| | Pf(Ec) | 9.08 | 0.40 | 0.56 |

Figure 1. Directed acyclic graph associated to the Bayesian model. Circles: random variables; squares: constants (initial parameters of the distributions of the variables); arrows: conditional dependence. Observed data (Obs. j-1) of *E. coli spp.* (EcO1, EcO2, and Ec), and *P. fluorescens* (Pf) with their growth rates µ distributed with a Normal distribution of mean m and standard deviation S. The standard deviation of errors ($\sigma_j$) for every microorganism is distributed with a Gamma distribution with parameters α. The combinations of the microorganisms in the single cultures and in the co-cultures give the observed data.

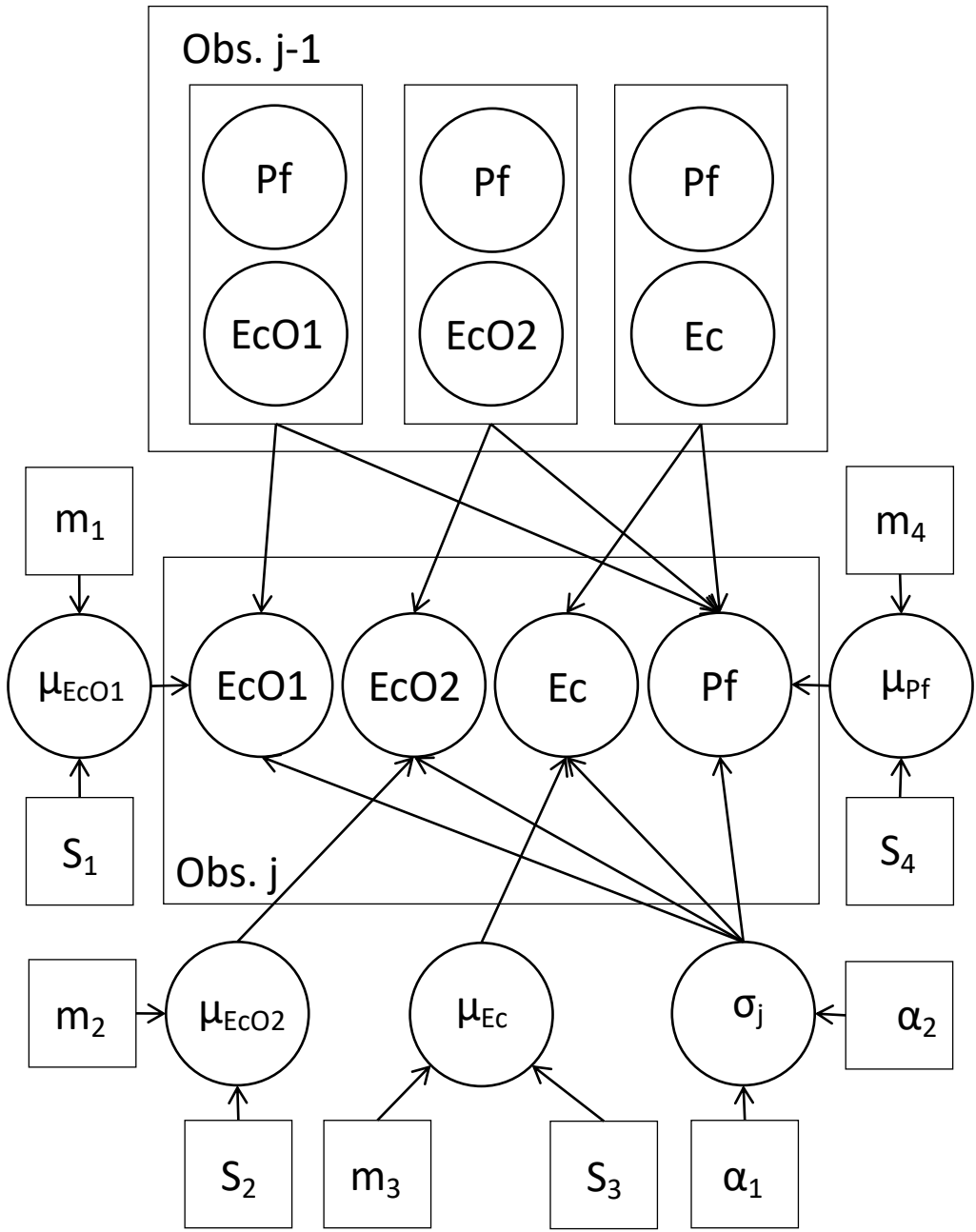

Figure 2. Bayesian inference of growth periods of co-cultures of *E. coli* spp. with *P. fluorescens* at 7 or 25°C in skimmed milk. Highest Posterior Density (HPD) 95% intervals (2.5 and 97.5%) are shown.

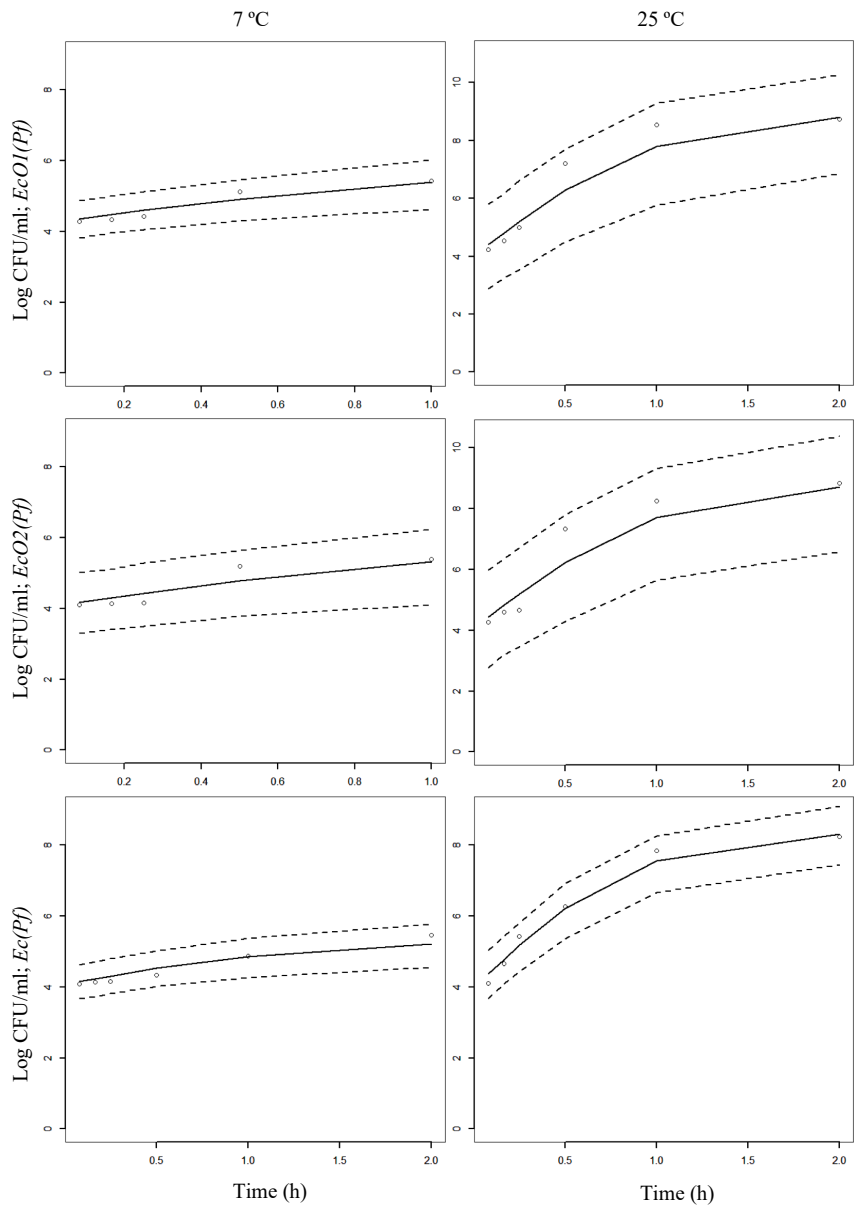

Figure 3. Bayesian inference of decline periods of co-cultures of *E. coli* spp. with *P. fluorescens* at 7 or 25°C in skimmed milk. Highest Posterior Density (HPD) 95% intervals (2.5 and 97.5%) are shown.

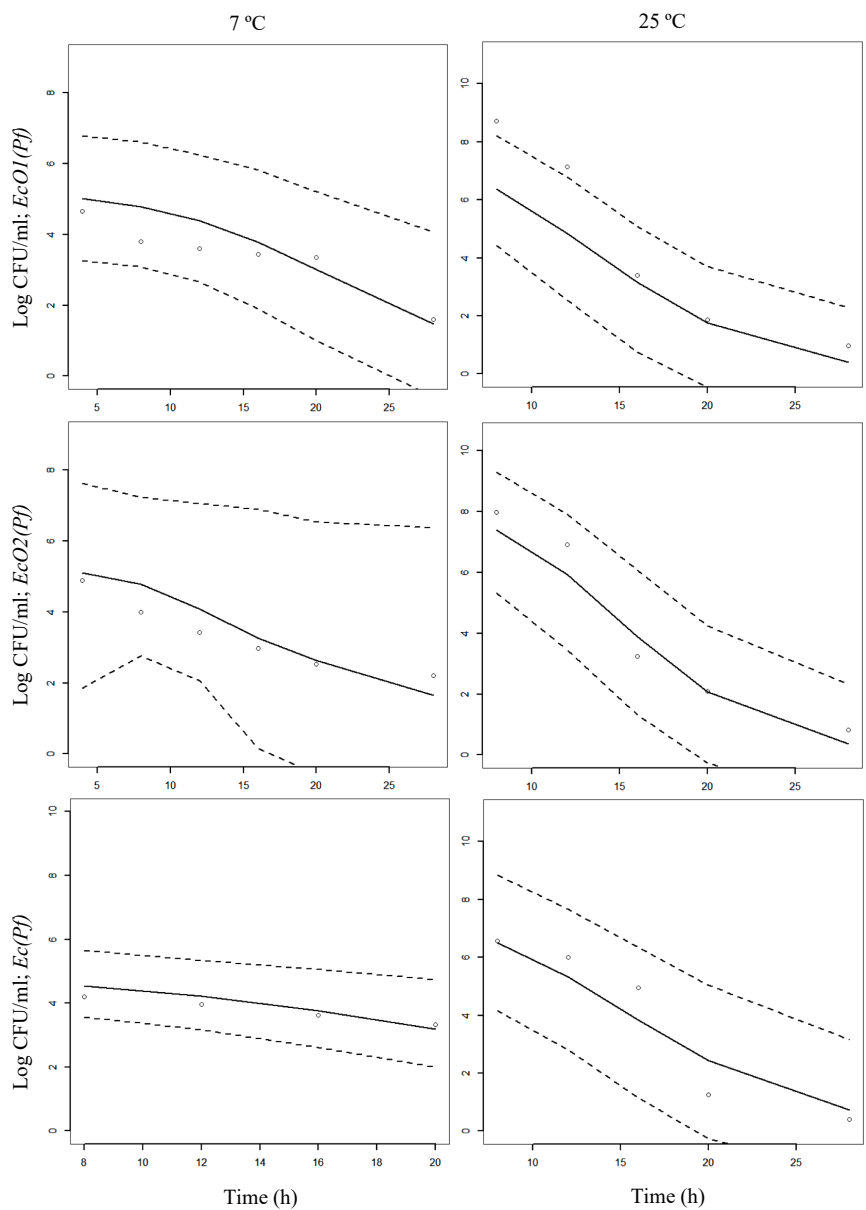

Figure 4. (A) Mosaic plot of the growth rates (μ mean) of *E. coli* spp and *P. fluorescens* cultured alone or in co-culture at 7, 13, 19 or 25ºC; vertical lines in the co-cultures columns (see cultures axis, i.e., Ec+Pf, EcO1+Pf, or EcO2+Pf) represent the strains not included on them; arrows indicate the data columns used for Figure B. (B) Scatter plot of μ values from *E. coli* spp. co-cultured with *P. fluorescens* (EcO1(Pf), EcO2(Pf), and Ec(Pf)) or single-cultured (EcO1, EcO2, and Ec).

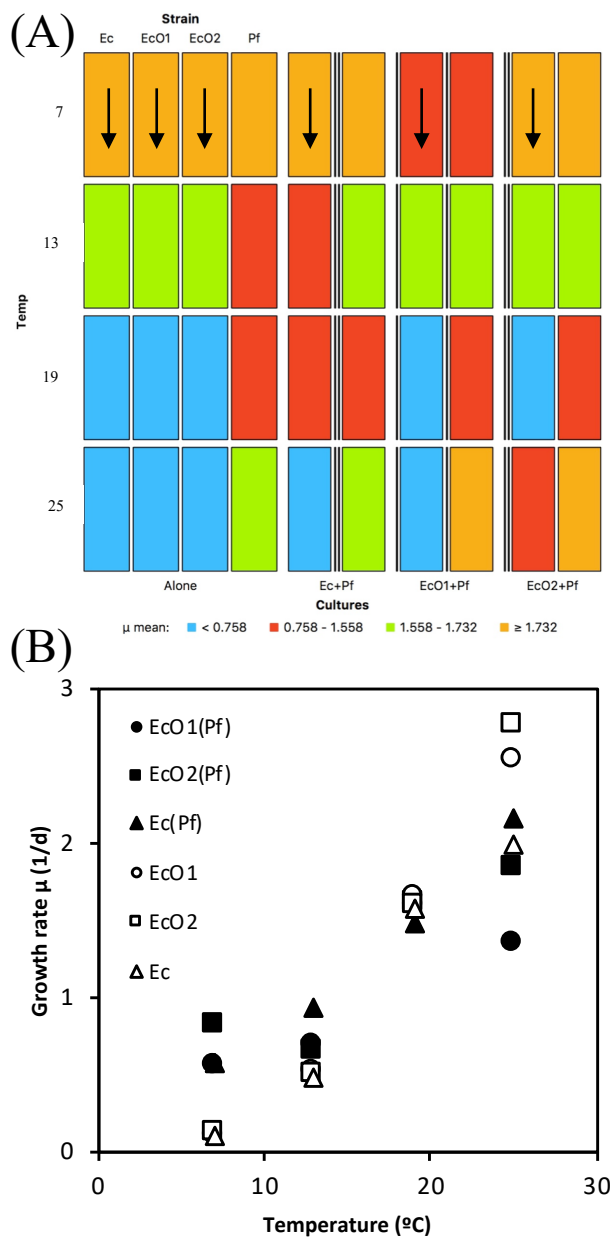

Figure S1. Hamiltonian Monte Carlo Method (HMCM) diagnosis plots of the growth rates of *E. coli* O157:H7 LCDC 86-51 (EcO1), *E. coli* O157:H7 ATCC 35150 (EcO2) or *E. coli* ATCC 8739 (Ec) co-cultured with *P. fluorescens* (EcO1(Pf), EcO2(Pf), or Ec(Pf)) at 7 °C. Panels show three plots for the growth rate parameter: traces or value estimated in each step of the HMM (left); the parameter posterior distributions (middle); and the autocorrelation functions for the parameter estimates (right). mu[1] and mu[2] are the growth rates of *E. coli* spp or *P. fluorescens*, respectively.

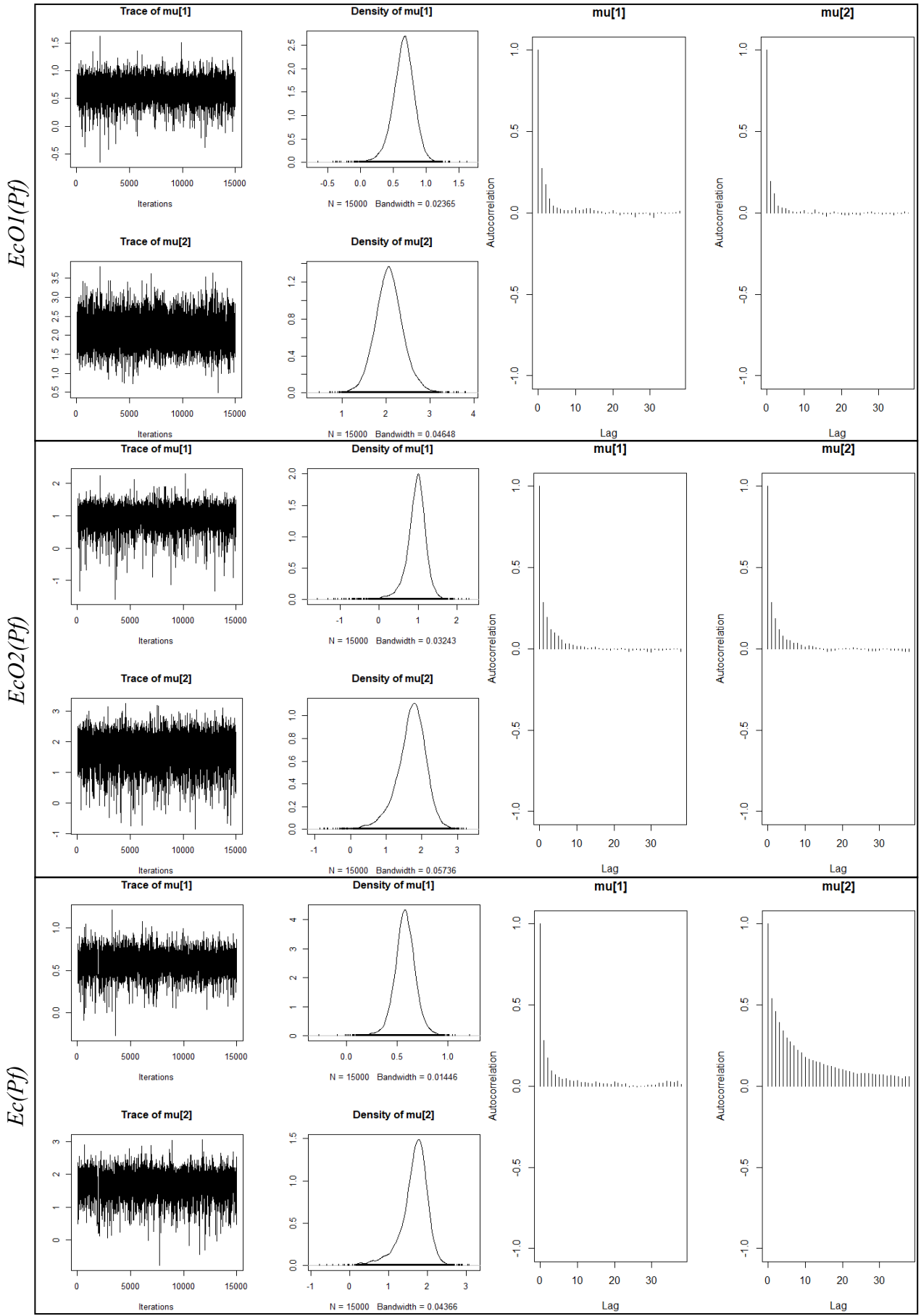

Figure S2. Hamiltonian Monte Carlo Method (HMCM) diagnosis plots of the growth rates of *E. coli* O157:H7 LCDC 86-51 (EcO1), *E. coli* O157:H7 ATCC 35150 (EcO2) or *E. coli* ATCC 8739 (Ec) co-cultured with *P. fluorescens* at 25 °C. See legends and explanations in Figure S1.

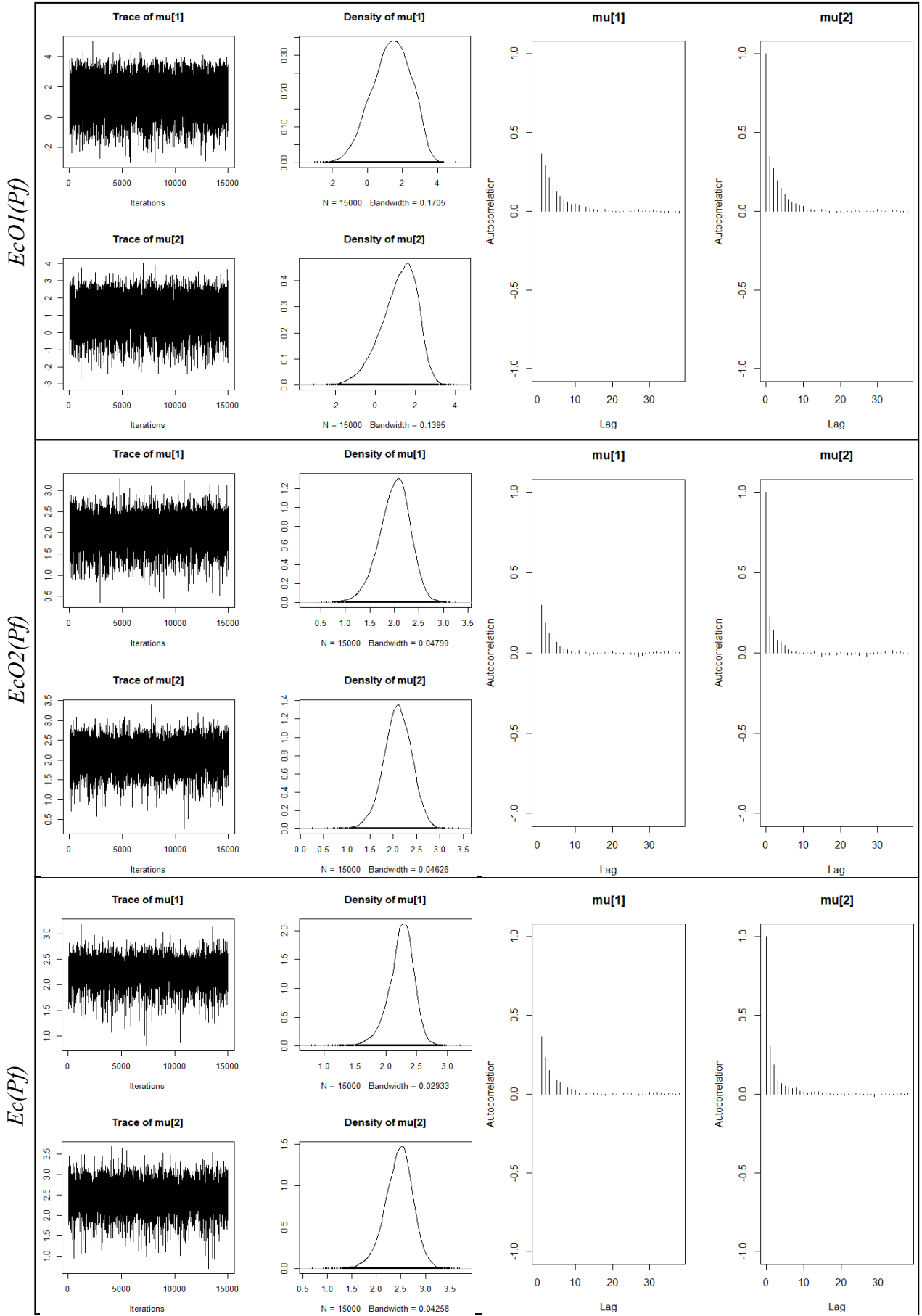

Figure S3. Hamiltonian Monte Carlo Method (HMCM) diagnosis plots of the decline rates of *E. coli* O157:H7 LCDC 86-51 (EcO1), *E. coli* O157:H7 ATCC 35150 (EcO2) or *E. coli* ATCC 8739 (Ec) co-cultured with *P. fluorescens* at 7 °C. See legends and explanations in Figure S1. k[1] and k[2] are the decline rates of *E. coli* spp or *P. fluorescens*, respectively.

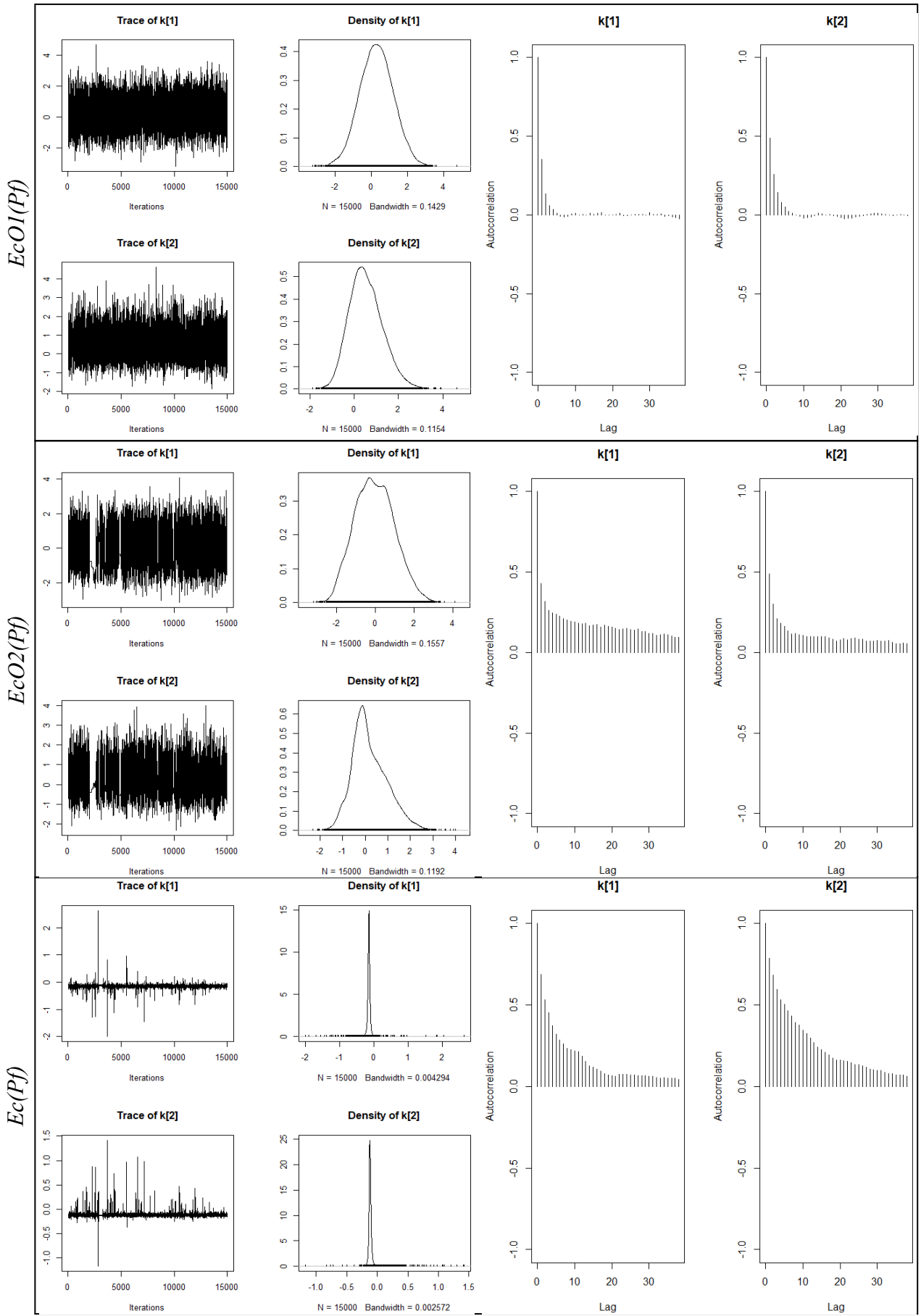

Figure S4. Hamiltonian Monte Carlo Method (HMCM) diagnosis plots of the decline rates of *E. coli* O157:H7 LCDC 86-51 (EcO1), *E. coli* O157:H7 ATCC 35150 (EcO2) or *E. coli* ATCC 8739 (Ec) co-cultured with *P. fluorescens* at 25 °C. See legends and explanations in Figures S1 and S3.

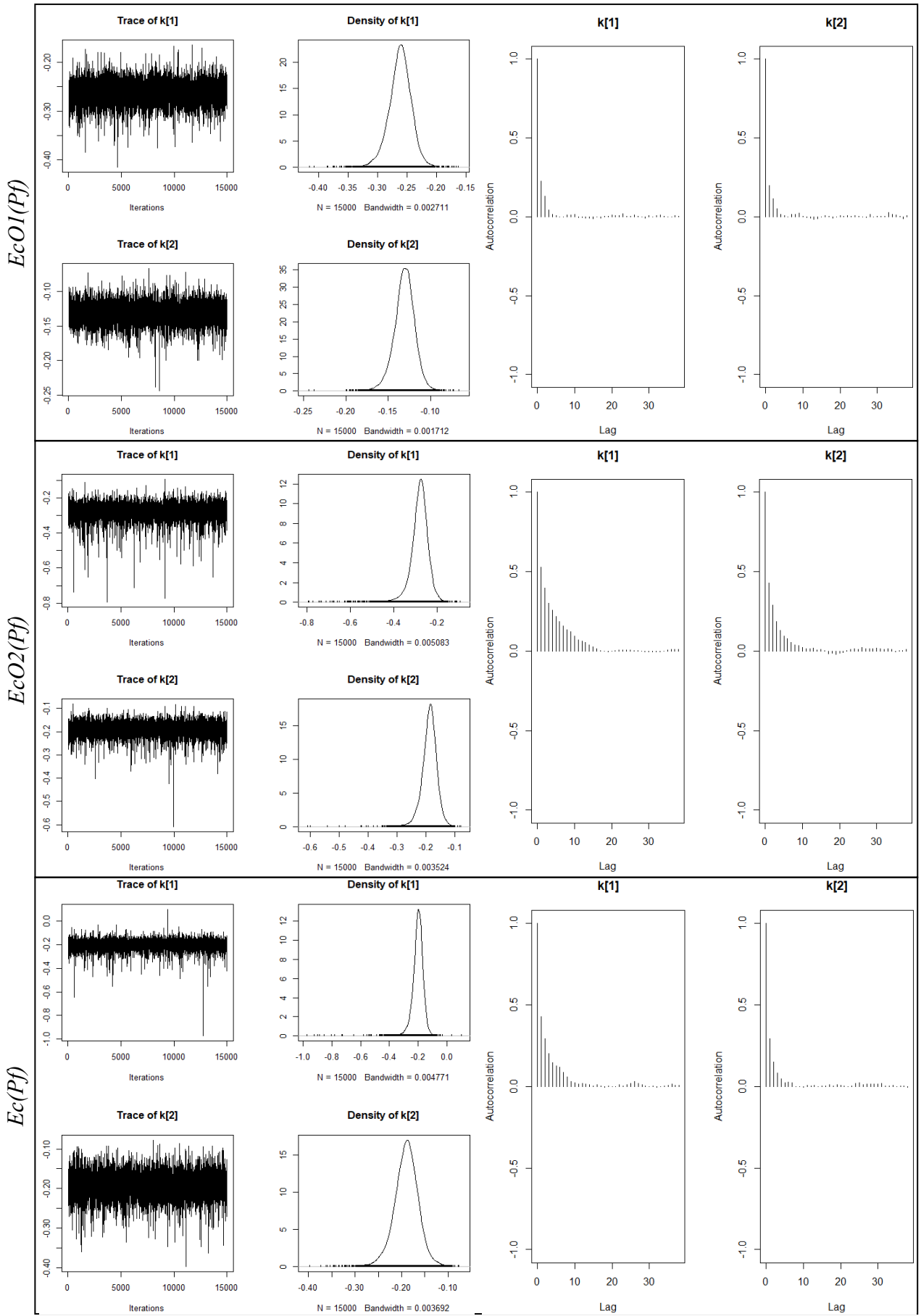